\documentstyle [12pt,a4p,epsfig,amsmath,multicol]{article}
\begin{document}
\vspace*{0.6cm}

\begin{center} 
{\normalsize\bf Translational invariance and the space-time Lorentz
 transformation with arbitrary spatial coordinates}
\end{center}
\vspace*{0.6cm}
\centerline{\footnotesize J.H.Field}
\baselineskip=13pt
\centerline{\footnotesize\it D\'{e}partement de Physique Nucl\'{e}aire et 
 Corpusculaire, Universit\'{e} de Gen\`{e}ve}
\baselineskip=12pt
\centerline{\footnotesize\it 24, quai Ernest-Ansermet CH-1211Gen\`{e}ve 4. }
\centerline{\footnotesize E-mail: john.field@cern.ch}
\baselineskip=13pt
\vspace*{0.9cm}
\abstract{Translational invariance requires that physical predictions
 are independent of the choice of spatial coordinate system used. The
  time dilatation effect of special relativity is shown to manifestly
  respect this invariance. Consideration of the space-time Lorentz
  transformation with arbitrary spatial coordinates shows that the spurious
  `length contraction' and `relativity of simultaneity' effects ---the latter
  violating translational invariance--- result from the use of a different
  spatial coordinate system to describe each of two spatially separated clocks at rest in a
  common inertial frame.}
\vspace*{0.9cm}
\normalsize\baselineskip=15pt
\setcounter{footnote}{0}
\renewcommand{\thefootnote}{\alph{footnote}}

  Translational invariance is a mathematical expression of the homogeneity of physical
  free space --the result of an experiment governed only by internal conditions does not
   depend on where, in space, it is performed. A corollary is that the prediction of
   the result of any such experiment does not depend on the choice of spatial
   coordinates used for its physical description. This is because moving the experiment to
   a different spatial position is mathematically equivalent to a change of the origins
   of coordinate axes $\vec{x} \rightarrow \vec{x}-\vec{x}_0$.
   In this letter, it is demonstrated that the space-time Lorentz transformation (LT) ---when
    correctly interpreted--- respects translational invariance, as just defined. As will
   be explained below, this is not the case in the conventional Einsteinian~\cite{Ein1}
     interpretation of the transformation.
    \par It is instructive to first discuss the space transformation equation in the 
      context of Galilean relativity. With a particular choice of coordinate axes, the
       Galilean space transformation for an object at rest in the inertial frame S', as
     observed in another such frame S, is:
     \begin{equation}
         x' = x - vt = 0.
     \end{equation}
      This equation describes an object at rest at the origin of S' that moves with
     uniform velocity, $v$, along the +ve $x$-axis in S. It is assumed that there
     is an array of synchronised clocks at rest in S and that $t$ is the time 
    recorded by any such clock. The spatial coordinate system in S is chosen so that
    $x = 0$ when $t = 0$. Introducing a more explicit notation and arbitrary
    coordinate origins in S and S', Eq.~(1) generalises to:
   \begin{equation}
         x'(t)- x'(0) = x(t) -x(0)- vt = 0.
     \end{equation}
     Thus $ x'(t)= x'(0)$ for all values of $t$, $ x(t)= x(0)$ for $t = 0$ and all
     values of $v$,  while the equation of motion
   of the object in S is given by the last member of (2).
    The `transformation equation' (2) is actually two separate and independent equations:
     \begin{eqnarray}
      x'(t) & = & x'(0), \\  
  x(t) & = & x(0)+vt. 
       \end{eqnarray}
 The spatial LT corresponding to Eq.~(2) is:
  \begin{equation}
         x'(t)- x'(0) = \gamma [x(t) -x(0)- vt] = 0.
     \end{equation}
     This, like (2), is equivalent to (3) and (4).
     In fact the multiplicative factor: $\gamma \equiv 1/\sqrt{1-(v/c)^2}$ in (5) may be replaced
     by a finite constant or an arbitrary finite function of $v/c$ and the transformation will still
     be equivalent to (3)
     and (4). The spatial description of the moving object is therefore identical for the Galilean
     and Lorentz transformations. Note that the constants $x'(0)$ and $x(0)$ that specify the positions
     of the origins of coordinates in the frames S' and S respectively {\it characterise a particular
     choice of coordinate systems}. In order to correctly calculate the spatial separations
     of objects at different positions it is essential to use the same coordinate system
     (i.e the same values of  $x'(0)$ and $x(0)$) to specify the positions of {\it all} the objects.
     As will be seen shortly, it is the failure to do this that lies at the origin of the
     spurious `length contraction' effect. 
      \par The temporal LT corresponding to (5) is:
   \begin{equation}
         t'(t) = \gamma [t- \frac{v(x(t)-x(0))}{c^2}].
    \end{equation}       
    The physical meaning of $t'(t)$ is the time recorded by a local clock at the position of the
    object in S' as observed, at time $t$, in the frame S. Because both $t$ and $x(t)$ appear
  on the right side of (6) one may think that $t'(t)$ depends on both $t$ and $x(t)$. This is
    Einstein's `relativity of simultaneity' (RS) effect. However, (4) may be used to eliminate
     $x(t)$ from the right side of (6) to yield the time dilatation (TD) formula first derived,
    in this way, directly from the LT, by Einstein~\cite{Ein1}:
  \begin{equation}
    t = \gamma t'(t)
 \end{equation}
   Since no spatial coordinates appear in this equation, it manifestly respects translational
   invariance. Also the clock in S' is `system externally synchronised'~\cite{MS} so that
     $t'(0) = 0$. 
    \par Keeping the same general coordinate system and `system external' synchronisation
     procedure as used in (5) and (6) consider now two clocks, C$_1$', C$_2$' at rest in S' at
       $x'_1(0)$ and $x'_2(0)$. If $t'_1(t)$ and  $t'_2(t)$ 
       are the observed times of the clocks in S at time $t$, then the relation
          (7) must hold for both of the clocks. In consequence,
   \begin{equation}
      \gamma t'_1(t) =  \gamma t'_2(t) = t
   \end{equation}
    so that  $t'_1(t) =  t'_2(t)$ ---there is no RS effect. Indeed this  follows from the
    the fact that $t'$  in Eq.~(7), contrary to Eq.~(6) without the additional condition
    (4), is a function only of $t$, not of $t$ and $x(t)$.
  \par Choosing  $x_1(0) = x'_1(0) = 0$ and $x_2(0) = L$, $x'_2(0) = L'$ the LT describing
  the transformation of points on the world lines of  C$_1$' and C$_2$' are:
     \begin{eqnarray}
        x'_1 & = & \gamma [x_1- v t_1] = 0, \\
        t'_1 & = & \gamma [t_1- \frac{v x_1}{c^2}], \\
        x'_2 - L' & = & \gamma [x_2-L- v t_2] = 0, \\
        t'_2 & = & \gamma [t_2- \frac{v(x_2-L)}{c^2}].
        \end{eqnarray}
   Consider simultaneous events in S': $ t'_1 =  t'_2 = t'$. Combining (9),(10) and (11),(12)
   gives 
      \begin{eqnarray}
       x_1(\beta)& = & v t_1 = \gamma \beta c t',  \\
                t_1(\beta) & = & \gamma t',   \\
  x_2(\beta)- L & = & v t_2 = \gamma \beta c t',  \\
                t_2(\beta) & = & \gamma t'
  \end{eqnarray}
   where the $\beta$ dependence of the coordinates is explicitly indicated.
   The identity: $\gamma^2 -\gamma^2 \beta^2 \equiv 1$ shows that (13),(14) and (15),(16) 
  are parametric formulae for the hyperbolae in $x-t$ space:
   \begin{equation}
    c^2  t_1(\beta)^2- x_1(\beta)^2 = c^2(t')^2 = c^2  t_2(\beta)^2- (x_2(\beta)-L)^2. 
   \end{equation}
     Since, from (14) and (16), $t_1(\beta) =  \gamma t' = t_2(\beta)$, (17) requires that
   \begin{equation}
   x_2(\beta) - x_1(\beta) = L.
  \end{equation}
     Because $L = x_2(0)$, is a constant independent of $\beta$, (18) holds for arbitrary
    values of the latter quantity. In particular, it holds when $\beta \rightarrow 0$,
     $x \rightarrow x'$, giving, in this case,  
   \begin{equation}
   x_2(0) - x_1(0) = x'_2 - x'_1 \equiv L' =  L.
  \end{equation}
    Thus the spatial separation of the clocks is a Lorentz-invariant quantity; 
     ---there is no relativistic `length contraction' (LC). 
  \par How the spurious and correlated RS and LC effects of conventional special relativity
   arise will now be explained. Following Einstein~\cite{Ein1} the choice $x(0) = x'(0) = 0$
    is made in the general LT (5) and (6) to give :
    \begin{eqnarray}
     x'(t) & = & \gamma[x(t) - v t] = 0,  \\
     t'(t) & = & \gamma[t - \frac{v x(t)}{c^2}].
     \end{eqnarray}
     Since the TD relation (7) does not depend on the choice of spatial coordinate
     system, it holds also when (20) and (21) are used. If the clock C$_1$' is placed
     at $x_1'(t) = x_1'(0) = 0$, it is, according to (20) and (21), synchronised so that $t'_1(t = 0) = 0$.
     The `LC effect' is derived by substituting the coordinates of the clock  C$_2$',
      placed at $x_2'(t) = x_2'(0) = L'$ in the first member of space transformation
      equation (20) and setting $t= 0$. This procedure is aleady in contradiction with the last member
     of (20) since it is assumed, in making this subsitution, that $L' \ne 0$. 
     After the above substitution in (20), and since  $x_1'(0) = x_1(0) = 0$,
      the following equation may be derived:
      \begin{equation}
      x_2'(0) - x_1'(0) = L' =  \gamma [x_2(0)-  x_1(0)] = \gamma L.
     \end{equation}
     This is the `LC effect'. However, the assumption on which (22) is based, $x_1'(0) = x_1(0) = 0$,
     is inconsistent with the formula from which (22) is derived:
    \begin{equation}
      x_2'(t) = L' =  \gamma [x_2(t)-  v t] \ne 0
     \end{equation}
     or
     \begin{equation}
     x_2'(t) - L' =  \gamma [x_2(t)-\frac{ L'}{\gamma} - v t] = 0
     \end{equation}
      Comparing (24) with the general formula (5) requires that
    \begin{eqnarray}
      x'_2(0) & = &  x'(0) = L',  \\
       x_2(0) & = &  x(0) = \frac{L'}{\gamma} =  \frac{x'(0)}{\gamma}.
    \end{eqnarray}
    These equations imply that in (23) a different coordinate system is used in the frame S to 
    specifiy the position of C$_2$' to the one used to specify the position
    of C$_1$' in Eqs.~(20) and (21) where $x'_1(0) =  x_1(0) = 0$. In fact, with the
   coordinate system corresponding to (23), $ x_2(0) = L'/\gamma$ and it follows that
    $x_1(0)$ is not zero but rather
     \[  x_1(0) = x_2(0)-L = L'/\gamma-L. \]
     The LT for C$_1$' and  C$_2$', using the coordinate
     systems defined by (25) and (26), are therefore:
   \begin{eqnarray}
 x'_1(t) & = & \gamma [ x_1(t)- L'/\gamma + L-v t] = 0, \\
  t'_1(t) & = & \gamma [t - \frac{v( x_1(t)- L'/\gamma + L)}{c^2}], \\
 x'_2(t)-L' & = & \gamma [ x_2(t)- L'/\gamma -v t] = 0, \\
  t'_2(t) & = & \gamma [t - \frac{v( x_2(t)- L'/\gamma)}{c^2}].
    \end{eqnarray}
  Eqs.~(27) and (29) give
   \begin{eqnarray}
      x_1(t) & = & L'/\gamma - L+v t,  \\
  x_2(t) & = & L'/\gamma+v t
   \end{eqnarray}
   so that
 \begin{equation}
 x_2(t)- x_1(t) =  L.
 \end{equation}
  Transposing (27) gives
  \begin{equation}
    x'_1(t)+L'  =  \gamma [ x_1(t) + L-v t].
  \end{equation}
  Taking the limit $v \rightarrow 0$, $ \gamma \rightarrow 1$, $x \rightarrow x'$ in (34)
   gives
 \begin{equation}
    x'_1(t)+L'  =  x'_1(t) + L
  \end{equation}
  so that $L = L'$, as in (19) above, and in contradiction with the `LC effect' of (22).  
 It can be seen that the spurious `LC effect' of (22) is a consequence of
   using different coordinate systems in the frame S to describe the two clocks:
   $ x_1(0) =  x'_1(0) = 0$ for C$_1$' and $x_2(0) = x'_2(0)/\gamma  =  L'/\gamma$
   for  C$_2$'. When the latter system is used for both clocks,
   as in (27)-(30), the LC effect does not occur.
   \par If it is incorrectly assumed that  $x(0) = 0$ in (6), applied
    to the clock  C$_2$', when the condition (23) is satisfied, it is found,
    instead of (30) that
 \begin{equation}
  t'_2(t, L')  =  \frac{t}{\gamma} - \frac{v L'}{c^2} \ne  t'_1(t) =  \frac{t}{\gamma}
 \end{equation}
   in contradiction to Eq.~(8) above, and violating
   translational invariance The spurious dependence of $t'_2$ on
   $L'$ in Eq.~(36) is Einstein's RS effect.
  \par Similar conclusions to those of the present letter have been obtained
   elsewhere, by a careful study of clock synchronisation procedures
   ~\cite{JHFLLT,JHFCRCS}. 
   \par To date, there is no experimental verification of the RS or LC effects of conventional
    special relativity theory~\cite{JHFLLT}, which are claimed to be spurious in the present
   letter. However the existence (or not) of the O($v/c$) RS effect of Eq.~(34) is easily
    tested using modern experimental techniques. Two experiments, using satellites
    in low Earth orbit, have been proposed to perform such a test~\cite{JHFRSE}.

\end{document}